\documentclass[12pt]{article}

\usepackage{scicite}

\usepackage{times}
\usepackage{psfig}

\newenvironment{sciabstract}{%
\begin{quote} \bf}
{\end{quote}}

\newcounter{lastnote}
\newenvironment{scilastnote}{%
\setcounter{lastnote}{\value{enumiv}}%
\addtocounter{lastnote}{+1}%
\begin{list}%
{\arabic{lastnote}.}
{\setlength{\leftmargin}{.22in}}
{\setlength{\labelsep}{.5em}}}
{\end{list}}

\title{A bright millisecond radio burst of extragalactic origin}

\author
{D.~R.~Lorimer,$^{1,2\ast}$ M.~Bailes,$^3$  M.~A.~McLaughlin,$^{1,2}$\\
  D.~J.~Narkevic,$^1$ F.~Crawford$^4$\\
\normalsize{$^{1}$Department of Physics, West Virginia University, P.O. Box 6315, WV 26506 USA}\\
\normalsize{$^{2}$National Radio Astronomy Observatory, P.O.~Box 2, Green Bank, WV 24944}\\
\normalsize{$^{3}$Centre for Astrophysics and Supercomputing, Swinburne University of Technology,}\\ \normalsize{P.O.~Box 218, Hawthorn, Vic, 3122, Australia}\\
\normalsize{$^{4}$Department of Physics and Astronomy, Franklin and Marshall College, Lancaster, PA 17604 USA}\\
\normalsize{$^\ast$To whom correspondence should be addressed; E-mail:  Duncan.Lorimer@mail.wvu.edu.}
}

\date{Accepted for publication in the journal Science}

\begin{document} 


\baselineskip24pt


\maketitle



\begin{sciabstract}
Pulsar surveys offer one of the few opportunities to monitor even a
small fraction ($\sim 10^{-5}$) of the radio sky for impulsive
burst-like events with millisecond durations. In analysis of archival
survey data, we have discovered a 30-Jy dispersed burst of duration
$<5$~ms located three degrees from the Small Magellanic Cloud. The
burst properties argue against a physical association with our Galaxy
or the Small Magellanic Cloud. Current models for the free electron
content in the Universe imply the burst is $<1$~Gpc distant. No
further bursts are seen in 90-hr of additional observations, implying
that it was a singular event such as a supernova or coalescence of
relativistic objects. Hundreds of similar events could occur every day
and act as insightful cosmological probes.
\end{sciabstract}


Transient radio sources are difficult to detect, but can potentially
provide insights into a wide variety of astrophysical
phenomena\cite{clm04}.  Of particular interest is the detection of
short-duration ($\leq$~few ms) radio bursts that may be produced by
exotic events at cosmological distances such as merging neutron
stars\cite{hl01b} or evaporating black holes\cite{ree77}.  Pulsar
surveys are currently one of the only records of the sky with good
sensitivity to radio bursts, and the necessary temporal and spectral
resolution required to unambiguously discriminate between
short-duration astrophysical bursts and terrestrial interference.
Indeed, they have recently been successfully mined to detect a new
Galactic population of transients associated with rotating neutron
stars\cite{mll+06}.  The burst we report here, however, has a
significantly higher inferred energy output than this class and is not
observed to repeat. This burst, therefore, represents an entirely new
phenomenon.

The burst was discovered during a search of archival data from a
1.4-GHz survey of the Magellanic Clouds\cite{mfl+06} using the
multibeam receiver on the 64-m Parkes radio telescope\cite{swb+96} in
Australia. The survey consisted of 209 telescope pointings, each
lasting 2.3~hr. During each pointing, the multibeam receiver collected
independent signals from 13 different positions (beams) on the sky.
The data from each beam were one-bit sampled every millisecond over 96
frequency channels spanning a 288-MHz wide band. Radio signals from
all celestial sources propagate through a cold ionized plasma of free
electrons before reaching the telescope. The plasma, which exists
within our Galaxy and in extragalactic space, has a refractive index
which is frequency dependent. As a result, any radio signal of
astrophysical origin should exhibit a quadratic shift in their arrival
time as a function of frequency, with the only unknown being the
integrated column density of free electrons along the line of sight,
known as the dispersion measure (DM).  Full details of the data
reduction procedure to account for this effect, and to search for
individual dispersed bursts, are given in the supporting online
material. In brief, for each beam, the effects of interstellar
dispersion were minimized for 183 trial DMs in the range
0--500~cm$^{-3}$~pc.  The data were then searched for individual
pulses with signal-to-noise ratios (S/N) greater than four using a
matched filtering technique\cite{cm03} optimized for pulse widths in
the range 1~ms--1~s. The burst was detected in data taken on 2001
August 24 with DM~$=375$~cm$^{-3}$~pc contemporaneously in three
neighboring beams (Fig.~1), and is located approximately three degrees
south of the center of the Small Magellanic Cloud (SMC).

\nocite{gmrv01,ssd+99}

The pulse exhibits the characteristic quadratic delay as a function of
radio frequency (Fig.~2) expected from dispersion by a cold ionized
plasma along the line of sight\cite{lk05}. Also evident is a
significant evolution of pulse width across the observing frequency
band.  The behavior we observe, where the pulse width $W$ scales with
frequency $f$ as $W \propto f^{-4.8\pm 0.4}$, is consistent with
pulse-width evolution due to interstellar scattering with a Kolmogorov
power law ($W \propto f^{-4}$, 11)\nocite{lj76}.  The filterbank
system has finite frequency and time resolution, which effectively
sets an upper limit to the intrinsic pulse width $W_{\rm int} =
5$~ms. We represent this below by the parameter $W_5 = W_{\rm
int}/5$~ms. Note that it is entirely possible that the intrinsic width
could be much smaller than observed (i.e.~$W_5 \ll 1$) and the
combination of intergalactic scattering and our instrumentation smear
it to the resolution we see in Fig.~2.

We can estimate the flux density of the radio burst in two ways.  For
the strongest detection, which saturated the single-bit digitizer in
the observing system, we make use of the fact that the integrating
circuit that sets the mean levels and thresholds is analog. When
exposed to a source of strength comparable to the system equivalent
flux density, an absorption feature in the profile is induced that can
be used to estimate the integrated burst energy. For a 5-ms burst we
estimate the peak flux to be 40~Jy (1~Jy~$\equiv
10^{-26}$~W~m$^{-2}$~Hz$^{-1}$).  Using the detections from the
neighboring beam positions, and the measured response of the multibeam
system as a function of off-axis position\cite{swb+96}, we determined
the peak flux density to be at least 20~Jy. We therefore adopt a burst
flux of $30 \pm 10$~Jy, which is consistent with our measurements, for
the remaining discussion.  Although we have only a limited information
on the flux density spectrum, as seen in Fig.~2, the pulse intensity
increases at the lowest frequencies of our observing band. This
implies that the flux density $S$ scales with observing frequency $f$
as $S \propto f^{-4}$.

It is very difficult to attribute this burst to anything but a
celestial source. The frequency dispersion and pulse-width frequency
evolution argue for a cosmic origin. It is very unlikely that a
swept-frequency transmitter could mimic both the cold plasma
dispersion law to high accuracy (see Fig.~2), and have a scattering
relation consistent with the Kolmogorov power law. Furthermore,
terrestrial interference often repeats, and this was the only
significantly dispersed burst detected with S/N~$>10$ in the analysis
of data from almost 3000 separate positions.  Sources with flux
densities greater than about 1 Jy are typically detected in multiple
receivers of the multibeam system.  While this is true for both
terrestrial and astrophysical sources, the telescope had an elevation
of $\sim 60^{\circ}$ at the time of the observation, making it
virtually impossible for ground-based transmitters to be responsible
for a source that was only detected in three adjacent beams of the
pointing.

We have extensively searched for subsequent radio pulses from this
enigmatic source. Including the original detection, there were a total
of 27 beams in the survey data which pointed within 30 arcmin of the
nominal burst position. These observations, which total 50 hr, were
carried out between 2001 June 19 and July 24 and show no significant
bursts.  In April 2007 we carried out 40~hr of follow-up observations
with the Parkes telescope at 1.4~GHz with similar sensitivity to the
original observation. No bursts were found in a search over the DM
range 0--500~cm$^{-3}$~pc. These dedicated follow-up observations
provide the best sensitivity and imply that the event rate must be
less than 0.025~hr$^{-1}$ for bursts with S/N~$>6$, i.e.~a 1.4-GHz
peak flux density greater than 300~mJy.  The data were also searched
for periodic radio signals using standard techniques\cite{lk05} with
null results.

The Galactic latitude ($b=-41.8^{\circ}$) and high DM of the burst
make it highly improbable for the source to be located within our
Galaxy. The most recent model of the Galactic distribution of free
electrons\cite{cl02} predicts a DM contribution of only
25~cm$^{-3}$~pc for this line of sight. In fact, of over 1700 pulsars
currently known, none of the 730 with $|b|>3.5^{\circ}$ has
DM~$>375$~cm$^{-3}$~pc. The DM is also far higher than any of the 18
known radio pulsars in the Magellanic clouds\cite{mfl+06}, the largest
of which is for PSR~J0131--7310 in the SMC with
DM~$=205$~cm$^{-3}$~pc. The other four known radio pulsars in the SMC
have DMs of 70, 76, 105 and 125~cm$^{-3}$~pc.  The high DM of
PSR~J0131--7310 is attributed\cite{mfl+06} to its location in an HII
region (see Fig.~1).  We have examined archival survey data to look
for ionized structure such as H$\alpha$ filaments or HII regions which
could similarly explain the anomalously large DM of the burst. No such
features are apparent. The source lies three degrees south from the
center of the SMC, placing it outside all known contours of radio,
infrared, optical and high-energy emission from the SMC. This, and the
high DM, strongly suggest that the source is well beyond the SMC,
which lies $61 \pm 3$~kpc away\cite{hhh05}.

No published gamma-ray burst or supernova explosion is known at this
epoch or position, and no significant gamma-ray events were detected
by the Third Interplanetary Network\cite{hmk+06,hur07} around the time
of the radio burst.  The Principal Galaxy Catalog (PGC;
16)\nocite{ppp+03} was searched for potential hosts to the burst
source. The nearest candidate (PGC~246336) is located 5 arcmin south
of the nominal burst position, but the non-detection of the burst in
the beam south of the brightest detection appears to rule out an
association. If the putative host galaxy were similar in type to the
Milky Way, the non detection in the PGC (limiting B magnitude of 18)
implies a rough lower limit of $\sim$~600~Mpc on the distance to the
source.

We can place an upper bound on the likely distance to the burst from
our DM measurement.  Assuming a homogeneous intergalactic medium in
which all baryons are fully ionized, the intergalactic DM is
expected\cite{iok03,ino04} to scale with redshift, $z$, as DM~$ \sim
1200 z$~cm$^{-3}$~pc for $z \leq 2$. Subtracting the expected
contribution to the DM from our Galaxy, we infer $z=0.3$ which
corresponds to a distance of about 1~Gpc. This is likely an upper
limit, since a host Galaxy and local environment could both contribute
to the observed DM. Using the electron density model for our
Galaxy\cite{cl02} as a guide, we estimate that there is a 25\%
probability that the DM contribution from a putative host galaxy is $>
100$~cm$^{-3}$~pc and, hence, $z<0.2$. Obviously the more distant the
source, the more powerful it becomes as a potential cosmological
probe. The sole event, however, offers little hope of a definitive
answer at this stage. To enable some indicative calculations about
potential source luminosity and event rates we adopt a distance of
500~Mpc. This corresponds to $z \sim 0.12$ and a host galaxy DM of
200~cm$^{-3}$~pc. In recognition of the considerable distance
uncertainty, we parameterize this as $D_{500}=D/500$~Mpc.  If this
source is well beyond the local group, it would provide the first
definitive limit on the ionized column density of the intra-cluster
medium, which is currently poorly constrained \cite{mb99}.

What is the nature of the burst source?  From the observed burst
duration, flux density and distance, we estimate the brightness
temperature and energy released to be $\sim 10^{34} (D_{500}/W_5)^2$~K
and $\sim 10^{33} W_5 D_{500}^2$~J respectively. These, and
light-travel-time arguments which limit the source size to $<1500$~km,
for a non-relativistic source imply a coherent emission process from a
compact region.  Relativistic sources with bulk velocity $v$ are
larger by a factor of either $\Gamma$ (for a steady jet model) or
$\Gamma^2$ (for an impulsive blast model), where the Lorentz factor
$\Gamma = (1-v^2/c^2)^{-1/2}$ and $c$ is the speed of light.

The only two currently known radio sources capable of producing such
bursts are the rotating radio transients (RRATs), thought to be
produced by intermittent pulsars \cite{mll+06}, and giant pulses from
either a millisecond pulsar or young energetic pulsar. A typical pulse
from a RRAT would only be detectable out to $\sim 6$~kpc with our
observing system. Even some of the brightest giant pulses from the
Crab pulsar, with peak luminosities of 4~kJy~kpc$^2$\cite{cbh+04},
would be observable out to $\sim 100$~kpc with the same system. In
addition, both the RRAT bursts and giant pulses follow power-law
distributions of pulse energies. The strength of this burst, which is
some two orders of magnitude above our detection threshold, should
have easily led to many events at lower pulse energies, either in the
original survey data or follow-up observations.  Hence, it appears to
represent an entirely new class of radio source.

To estimate the rate of similar events in the radio sky, we note that
the survey we have analyzed observed for 20~d and was sensitive to
bursts of this intensity over an area of about 9~deg$^2$
(i.e.~$1/4500^{\rm th}$ of the entire sky) at any given time.
Assuming the bursts to be distributed isotropically over the sky, we
infer a nominal rate of $\sim 4500/20=225$ similar events per day.
Given our observing system parameters, we estimate that a $10^{33}$-Jy
radio burst would be detectable out to $z \sim 0.3$, or a distance of
1~Gpc.  The corresponding cosmological rate for bursts of this energy
is therefore $\sim 50$~d$^{-1}$~Gpc$^{-3}$.  Though considerably
uncertain, this is somewhat higher than the corresponding estimates of
other astrophysical sources, such as binary neutron star inspirals
($\sim 3$~d$^{-1}$~Gpc$^{-3}$; 21)\nocite{kkl+04} and gamma-ray bursts
($\sim 4$~d$^{-1}$~Gpc$^{-3}$; 22)\nocite{gd07}, but well below the
rate of core-collapse supernovae ($\sim 1000$~d$^{-1}$~Gpc$^{-3}$;
23)\nocite{mdp98}. Although the implied rate is compatible with
gamma-ray bursts, the brightness temperature and radio frequency we
observe for this burst are significantly higher than currently
discussed mechanisms or limitations for the observation of prompt
radio emission from these sources \cite{mac07}.

Regardless of the physical origin of this burst, we predict that
existing data from other pulsar surveys with the Parkes Multibeam
system\cite{mlc+01,ebsb01,bjd+06,jbo+07} should contain several
similar bursts.  Their discovery would permit a more reliable estimate
of the overall event rate. The only other published survey for radio
transients on this timescale \cite{alv89} did not have sufficient
sensitivity to detect similar events at the rate predicted here.  At
lower frequencies ($\sim 400$~MHz) where many pulsar surveys were
conducted, although the steep spectral index of the source implies an
even higher flux density, the predicted scattering time ($\sim 2$~s)
would make the bursts difficult to detect over the radiometer
noise. At frequencies near 100~MHz where low frequency arrays
currently under construction will operate\cite{svk+06}, the predicted
scattering time would be of order several minutes, and hence be
undetectable.

Perhaps the most intriguing feature of this burst is its 30-Jy
strength. While this has allowed us to make a convincing case for its
extraterrestrial nature, the fact that it is over 100 times our
detection threshold makes its uniqueness puzzling. Often, astronomical
sources have a flux distribution that would naturally lead to many
burst detections of lower significance; such events are not observed
in our data. If, on the other hand, this burst was a rare standard
candle, more distant sources would have such large DMs that they would
be both significantly red-shifted to lower radio frequencies and
outside our attempted dispersion trials. If redshifts of their host
galaxies are measurable, the potential of a population of radio bursts
at cosmological distances to probe the ionized intergalactic
medium\cite{gin73} is very exciting, especially given the construction
of wide-field instruments\cite{jef+07} in preparation for the Square
Kilometre Array\cite{wke+04}.



\begin{scilastnote}
\item The Parkes radio telescope is part of the Australia Telescope,
which is funded by the Commonwealth of Australia for operation as a
National Facility managed by CSIRO. We thank Dick Manchester for
making the archival data available to us. This research has made use
of data obtained from the High Energy Astrophysics Science Archive
Research Center (HEASARC), provided by NASA's Goddard Space Flight
Center. We thank Kevin Hurley for providing access to the GCN network
Archive and Vlad Kondratiev, Steven Tingay, Simon Johnston, Fernando
Camilo and Joss Bland-Hawthorn for useful comments on the
manuscript. We acknowledge the prompt awarding of follow-up time by
the ATNF Director and thank Lawrence Toomey and Peter Sullivan for
observing assistance.
\end{scilastnote}


\begin{figure}
\psfig{file=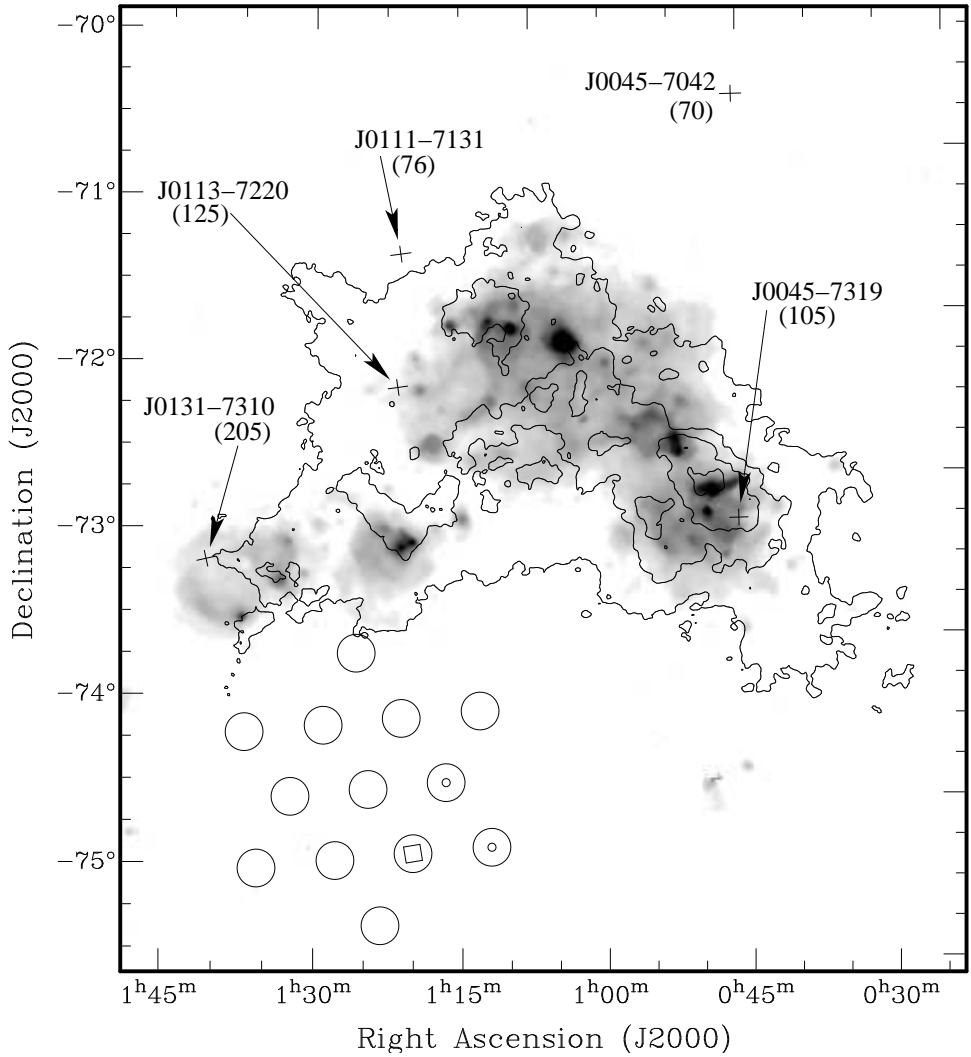,width=14cm}
\caption{
Multi-wavelength image of the field surrounding the burst. The gray
scale and contours respectively show H$\alpha$ and HI emission
associated with the SMC\cite{gmrv01,ssd+99}. Crosses mark the
positions of the five known radio pulsars in the SMC and are annotated
with their names and DMs in parentheses in units of cm$^{-3}$~pc.  The
open circles show the positions of each of the 13 beams in the survey
pointing of diameter equal to the half-power width. The strongest
detection saturated the single-bit digitizers in the data acquisition
system, indicating that its S/N~$\gg 23$. Its location is marked with
a square at right ascension $01^{\rm h}$ 18$^{\rm m}$ 06$^{\rm s}$ and
declination $-75^{\circ}$ 12$^{\prime}$ 19$^{\prime\prime}$ (J2000
coordinates).  The other two detections (with S/Ns of 14 and 21) are
marked with smaller circles. The saturation makes the true position
difficult to localize accurately. Based on the half-power width of the
multibeam system, the positional uncertainty is nominally $\pm
7'$. However, the true position is probably slightly ($\sim$~few
arcmin) north-west of this position given the non-detection of the
burst in the other beams.
}
\end{figure}

\begin{figure}
\psfig{file=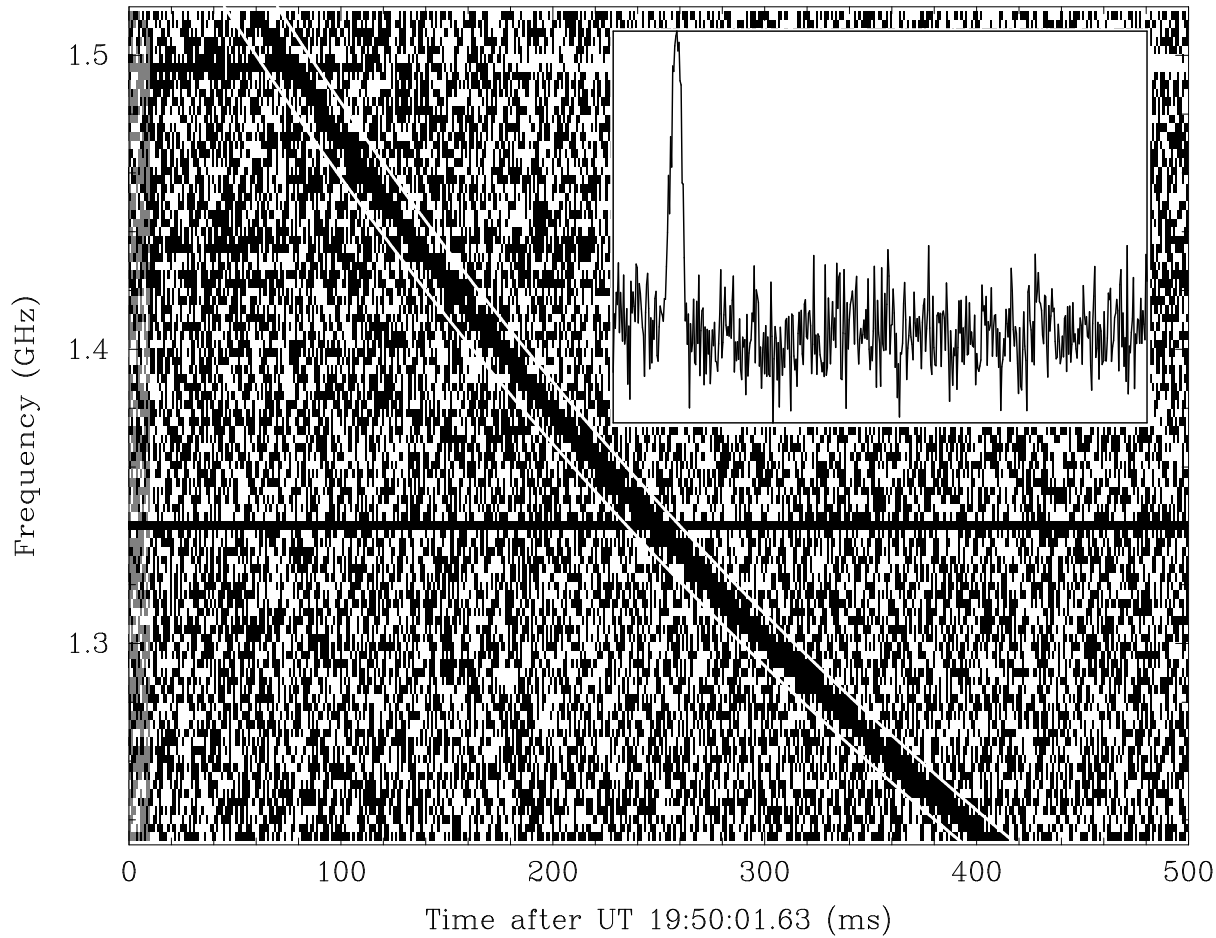,width=\textwidth}
\caption{
Frequency evolution and integrated pulse shape of the radio burst. The
survey data, collected on 2001 August 24, are shown here as a
two-dimensional `waterfall plot' of intensity as a function of radio
frequency versus time. The dispersion is clearly seen as a quadratic
sweep across the frequency band, with broadening towards lower
frequencies. From a measurement of the pulse delay across the receiver
band using standard pulsar timing techniques, we determine the DM to
be $375 \pm 1$~cm$^{-3}$~pc.  The two white lines separated by 15~ms
that bound the pulse show the expected behavior for the cold-plasma
dispersion law assuming a DM of 375 cm$^{-3}$~pc. The horizontal line
at $\sim 1.34$~GHz is an artifact in the data caused by a
malfunctioning frequency channel. This plot is for one of the offset
beams in which the digitizers were not saturated.  By splitting the
data into four frequency sub-bands we have measured both the
half-power pulse width and flux density spectrum over the observing
bandwidth. Accounting for pulse broadening due to known instrumental
effects, we determine a frequency scaling relationship for the
observed width $W = 4.6 \, {\rm ms} \, (f/1.4\,{\rm GHz})^{-4.8 \pm
0.4}$, where $f$ is the observing frequency. A power-law fit to the
mean flux densities obtained in each sub-band yields a spectral index
of $-4 \pm 1$. Inset: the total-power signal after a dispersive delay
correction assuming a DM of 375 cm$^{-3}$~pc and a reference frequency
of 1.5165~GHz. The time axis on the inner figure also spans the range
0--500~ms.
}
\end{figure}

\clearpage
\section*{Materials and methods}

We now describe the techniques used to search
for isolated dispersed bursts in the survey data. In general, we 
follow the methodology developed and described in detail by 
Cordes \& McLaughlin (S1).
The entire survey of the Magellanic clouds (S2)
amounted to 209 telescope pointings, each consisting of a 2.3-hr
observation of 13 independent positions (beams) on the sky.
The data for each beam were recorded as 96 frequency channels
spanning the band 1.2285--1.5165~GHz sampled contiguously at 1~ms intervals.
Each of the 2717 beams from the survey was processed independently using
freely available software (S3) in the following sequence:
interference excision,
time series generation, baseline removal, single-pulse detection
and diagnostic plot production.

\subsection*{Interference excision} 

To minimize the number of
impulsive bursts from terrestrial sources, for each sample 
the sum of the 96 frequency channels was computed. For
truly unbiased single-bit data, the expected sum is 48
(i.e.~half zeros and half ones) with a standard deviation
of $\sqrt{48}=6.9$. Those samples which deviated from
this ideal value by more than three standard deviations
were flagged and not used in any subsequent analysis.
This `clipping' process resulted in about 5\% of the
entire data being discarded.

\subsection*{Time series generation} 

Electromagnetic waves propagating
through an ionized plasma experience a frequency dependent delay 
across the observing
frequency band due to the dispersive effects of the plasma.
The difference in arrival times, $\Delta t$, 
between a pulse received at a high
frequency, $f_{\rm high}$, and a lower frequency, $f_{\rm low}$, is given
by the cold-plasma dispersion law (S4):
\begin{equation}
  \Delta t = 4.148808 \, {\rm ms} \, 
  \times 
     \left[
            \left(\frac{f_{\rm low}}{\rm GHz}\right)^{-2} -
            \left(\frac{f_{\rm high}}{\rm GHz}\right)^{-2} 
    \right] 
  \times \left(\frac{\rm DM}{{\rm cm}^{-3}\,{\rm pc}}\right),
\end{equation}
where DM is the integrated column density of free electrons in the
ionized medium. This formula was used to calculate the delays
of lower frequency channels with respect to the highest frequency
channel in our band. For a given DM, we can remove dispersion
by summing the individual frequency channels
together and appropriately delaying the lower-frequency channels with
respect to the highest frequency channel. The resulting data product
is known as a `dedispersed time series'. Since the DM of a
source is a-priori unknown, this process is repeated multiple times
to produce a set of time series spanning a range of DMs. In our analysis,
183 time series were produced for each beam, corresponding to 
the DM range 0--500~cm$^{-3}$~pc. The DM step size was chosen
such that the delay introduced at a slightly incorrect trial DM
was always less than one time sample. For this survey, any residual
broadening of pulses in the data is always less than 1~ms.

\subsection*{Time series normalization} 

The time series produced in the
previous step have an offset which reflects the system noise in the
receiver. This offset may vary significantly during the integration
due to receiver and background noise variations in the observing
system.  These effects were mitigated by dividing each time series
into 8 segments.  For each segment, we subtract its mean from each
sample therein and compute the resulting standard deviation,
$\sigma$. To ensure that the mean and standard deviation are not
biased by bright individual pulses, the procedure is performed twice,
with pulses detected during the first pass omitted from the mean and
standard deviation calculation of the second pass. 

\subsection*{Single-pulse detection} 

Individual samples are
considered potentially significant if they have amplitudes $A>4\sigma$
(S1). This simple thresholding process is most sensitive to pulses of width
equal to the sampling interval (i.e.~1~ms in our case) and
can be considered as an ideal matched filter to such pulses.
To optimize sensitivity to broader pulses, samples are
added in pairs, the standard deviation is recomputed and
the $4\sigma$ threshold is again applied. This process is
repeated a total of 10 times until a time resolution of 1.024~s
has been reached. 
The absolute time index, pulse width with the highest
signal-to-noise ratio (S/N; defined
simply as $A/\sigma$) and the S/N itself are stored for subsequent analysis.

The standard
deviation, $\sigma$, also contains useful information for calibration purposes.
Using a modified form of the radiometer equation (S4), the
root mean square noise fluctuations in Jy can be written as 
\begin{equation}
\Delta S_{\rm sys} =  
\frac{\beta T_{\rm sys}}{G\sqrt{n_{\rm p} N_{\rm add}
t_{\rm samp} 
\Delta f }} = C \sigma,
\end{equation}
where $\beta = \sqrt{\pi/2}$ is a factor accounting
for losses due to 1-bit digitization,
$T_{\rm sys} \approx 30$~K is the system noise temperature in the
receiver, $G \approx 0.7$~K~Jy$^{-1}$ is the antenna gain,
$N_{\rm add}$ is the number of times the time samples have been
added in pairs, $t_{\rm samp}=1$~ms is data sampling interval
and $\Delta f = 288$~MHz is the receiver bandwidth and
$C$ is the required scaling factor in units of
Jy per standard deviation.  In the flux density estimates given in the
paper, we calibrated the S/N into Jy by simply
multiplying them by $C$.

\subsection*{Diagnostic plotting} 

The search output is stored
for offline visual inspection as a set of diagnostic plots. We
show two examples of such plots for a detection of
the known pulsar in the Large Magellanic Cloud B0529$-$66
(Fig.~S1) and the discovery observation of the 
burst reported in this
paper (Fig.~S2). In both cases, the dispersed pulses are
clearly visible above the background noise.

Each diagnostic plot was carefully scrutinized and potentially
significant events were saved for follow-up analysis in which
the events were examined in the time--frequency plane as shown
in Fig.~2 of the main paper. Two known
pulsars B0529$-$66 and B0540$-$69 were identified as a result of this process.
The only remaining signal from the survey
which clearly follows the cold-plasma
dispersion law was the burst reported in the paper.

\section*{References and Notes}

\noindent
S1. J.~M. Cordes, M.~A. {McLaughlin}, {\it ApJ\/} {\bf 596}, 1142 (2003).

\bigskip
\noindent
S2. R.~N. {Manchester}, G.~{Fan}, A.~G. {Lyne}, V.~M. {Kaspi}, F.~{Crawford}, {\it
  ApJ\/} {\bf 649}, 235 (2006).

\bigskip
\noindent
S3. D.~R. Lorimer SIGPROC pulsar data analysis tools available online at
  http://sigproc.sourceforge.net.

\bigskip
\noindent
S4. D.~R. Lorimer, M.~Kramer, {\it {Handbook of Pulsar Astronomy}\/} (Cambridge
  University Press, 2005).

\clearpage
\begin{figure}
\psfig{file=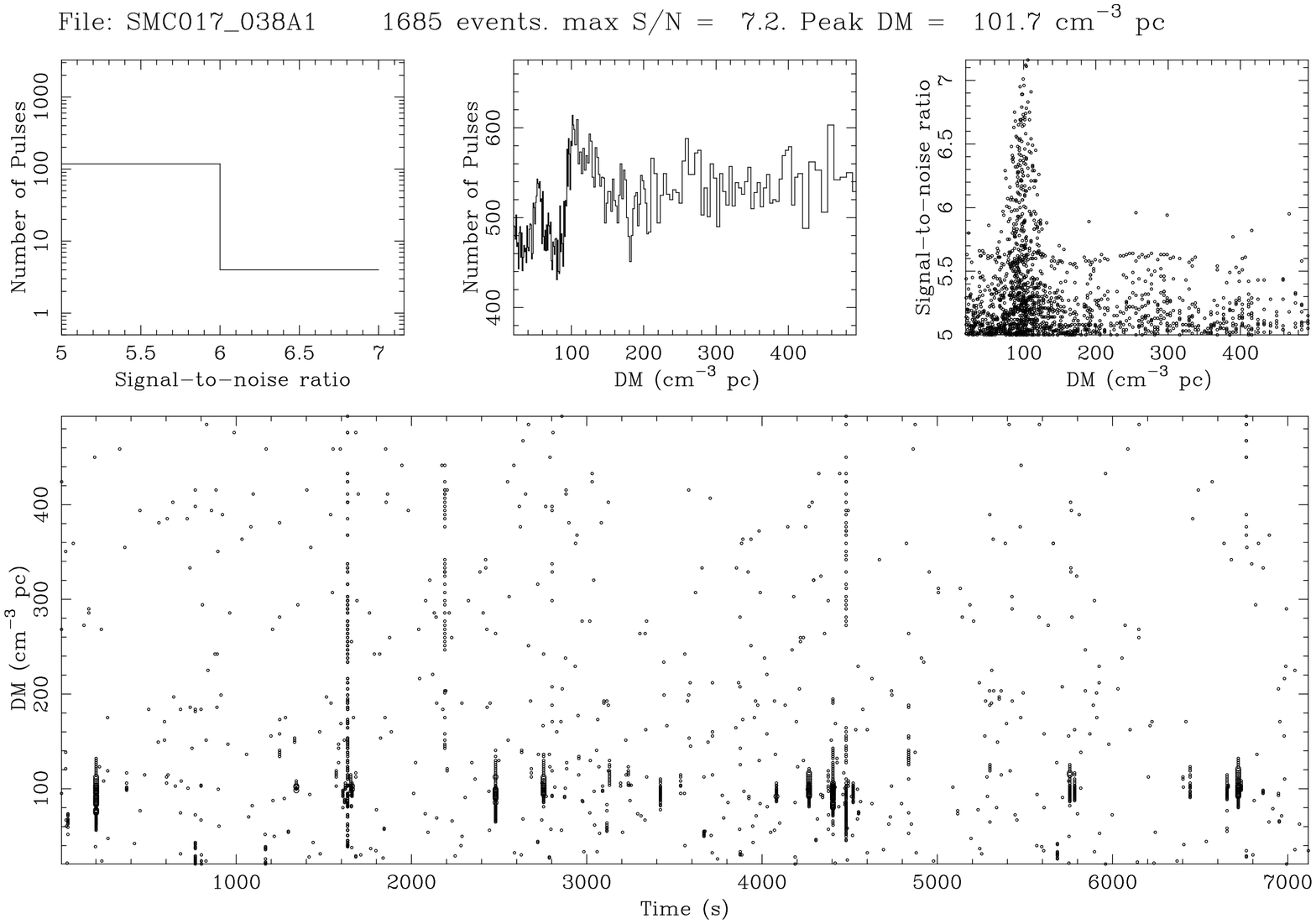,width=\textwidth}
\end{figure}

\noindent Figure S1: Single-pulse search output showing a survey
detection of the known pulsar B0529$-$66. From left to right, the
plots in the upper panel show the number of pulses detected as a
function of signal-to-noise ratio (S/N), the number as a function of
DM, and a scatter plot of S/N versus DM. The lower panel shows the
detected pulses as a function of observation time and DM, with the
size of the circles being proportional to S/N. The pulsar is clearly
visible as a band of occasional pulses in this diagram, with maximum
signal-to-noise at a DM of 101.7~cm$^{-3}$~pc, and also in the S/N
versus DM plot. Some locally generated interference detected across a
wider range of DMs is also present.

\clearpage
\begin{figure}
\psfig{file=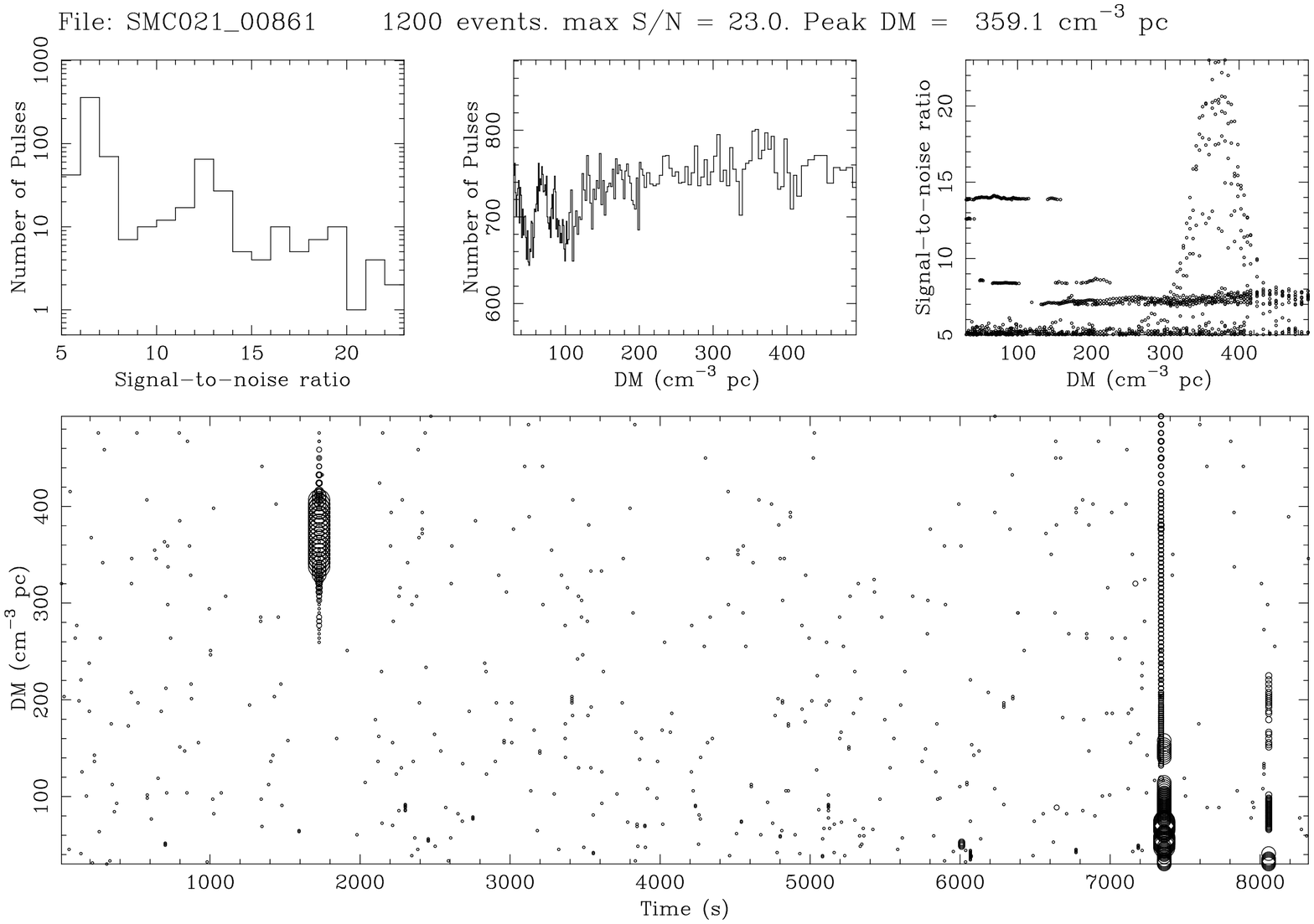,width=\textwidth}
\end{figure}

\noindent Figure S2: Single-pulse search output showing the discovery
of the burst reported in this paper. See caption to Fig.~S1 for
details of the diagnostic plots.  The burst is clearly visible in the
S/N versus DM plot with a S/N of 23 and DM of 360~cm$^{-3}$~pc (the
closest trial DM to the true value determined in the paper, which was
determined more precisely by an arrival-time analysis).  The event is
also seen in the lower panel as a highly dispersed isolated burst
occuring approximately 1650 s after the start of the observation. Note
also the presence of locally generated interference between 7000 and
8000~s.

\end{document}